\documentclass[12pt]{article}
\usepackage{amsmath}

\renewcommand \thesection{\arabic{section}.0}

\makeatletter
\newcounter {newsection}[section]
\renewcommand \thenewsection {\@arabic\c@section}
\renewcommand \thesubsection {\@arabic\c@newsection.\arabic{subsection}}
\newcounter {newnewsection}[section]
\renewcommand \thenewnewsection {\@Alph\c@section}
\makeatother

\def\bF{\mathbf{F}}
\def\bU{\mathbf{U}}
\def\bV{\mathbf{V}}
\def\bQ{\mathbf{Q}}
\def\bI{\mathbf{I}}

\def\nue{{\nu_e\to\nu_e}}
\def\lan{\langle}
\def\ran{\rangle}

\begin{document}


\begin{center}
{\Large Evidence for Internal Topological Constraints}\\
{\Large on Neutrino Mixtures}\\[3mm]
\hskip1in\vbox{\hbox{\large Gerald L. Fitzpatrick}
\hbox{\em PRI Research and Development Corp.}
\hbox{\em 12517 131 Ct. NE}
\hbox{\em Kirkland, WA 98034}
\hbox{glfitzpatrick@yahoo.com}}
\end{center}

\makeatletter
\renewcommand*\l@section[2]{%
  \ifnum \c@tocdepth >\z@
    \addpenalty\@secpenalty
    \addvspace{1.0em \@plus\p@}%
    \setlength\@tempdima{2.5em}%
    \begingroup
      \parindent \z@ \rightskip \@pnumwidth
      \parfillskip -\@pnumwidth
      \leavevmode \bfseries
      \advance\leftskip\@tempdima
      \hskip -\leftskip
      #1\nobreak\hfil \nobreak\hb@xt@\@pnumwidth{\hss #2}\par
    \endgroup
  \fi}
\makeatother

\begin{abstract}
\noindent 
A new internal description of fundamental fermions (quarks and leptons),
based on a matrix-generalization ($\bF$) of the scalar fermion-number $f$,  
predicts that only 
three families of quarks and leptons, and their associated neutrino flavors ($\nu_e$, 
$\nu_\mu$ and $\nu_\tau$), exist.  Moreover, this description appears to place important 
topological constraints on neutrino mixing. For example, with respect to $\bF$, 
the topology of the
$\nu_e$ ($\nu_\mu$ or $\nu_\tau$) neutrino flavor is found to be that of a cylinder (M\"obius strip). 
Assuming that  neutrino-neutrino transitions exhibiting topology change (i.e., $\nu_e\to\nu_\mu$, $\nu_\tau$, or $\nu_\mu, \nu_\tau\to\nu_e$) are \emph{suppressed},  while neutrino-neutrino
transitions without 
topology-change (i.e., $\nu_e\to \nu_e, \nu_\mu\to \nu_\mu, \nu_\tau\to\nu_\tau$ and $\nu_\mu\to\nu_\tau$ or $\nu_\tau\to\nu_\mu$) are relatively \emph{enhanced}, one may have an explanation for 
recent short-distance (i.e., ``atmospheric'')  observations of (nearly) maximal 
$\nu_\mu$-$\nu_\tau$
mixing.  To test this idea, I was able to 
use simple topological arguments to \emph{uniquely} determine the (symmetric) matrix of time-average probabilities
$M_{\alpha\beta}=\lan P_{\nu_\alpha\to\nu_\beta}\ran$ describing long-distance 
neutrino mixtures, which is \emph{identical} to that proposed by Georgi and 
Glashow on different grounds. In particular, using a conventional parameterization of the CKM-like neutrino mixing matrix $U$,
and the proposed topological constraints, I predict the following   mixing angle parameters: $s^2_2=c^2_2=s^2_3=c^2_3=\frac{1}{2}$ and $s^2_1=1$ or $c^2_1=0$, with 
a Dirac-type CP-noninvariant phase factor $e^{i\delta}$,  weakly constrained by $\sin^2\delta\ne 1$ or $\cos^2\delta\ne 0$ (i.e., CP violation is \emph{not} maximal). Then $M_{\alpha\beta}$ is predicted to be $M_{\alpha\beta}=(M_{11}=\frac{1}{2}, M_{12}=M_{13}=\frac{1}{4}, M_{22}=M_{23}=M_{33}=\frac{3}{8})$. Experimental verification of these predictions
would provide strong circumstantial evidence in support of the new description of fundamental fermions, which 
requires, among other things, that
the $\nu_e$ and 
($\nu_\mu$  or $\nu_\tau$) neutrino flavors
start and end ``life'' as \emph{topologically-distinct} quantum objects.
\end{abstract}
\newpage

\section{Introduction}

Except where explicitly prevented by some ``absolute'' conservation law
(e.g., the conservation of electric charge or spin angular momentum),
quantum mechanics generally permits transitions between states having
\emph{different} topologies [1]. While a change in topology may be energetically
(or otherwise) inhibited, unavoidable quantum fluctuations are expected to
\emph{catalyze} such processes. Hence, there is always the possibility of
\emph{mixing} between otherwise similar quantum states having \emph{distinct}
topologies [2].

Recently, a new internal description of fundamental fermions (quarks and 
leptons) was proposed [3].  The new description is based on a 
matrix-generalization $\bF$ of the scalar fermion number $f$.
One of the main predictions of the new description is that only three 
families of quarks and leptons exist---hence that there are only three low-mass 
neutrino flavors, namely, the $\nu_e$, $\nu_\mu$  and $\nu_\tau$ neutrinos.  
Moreover, because of 
the way fundamental fermions are represented by certain geometric objects in 
the space on which the matrix transformation $\bF$ acts (See Appendix A, and Ref.\ 3, p.\ 57,
pp. 85--87,  and Ref.\ 4, pp.\ 244--255),
the $\nu_e$ and ($\nu_\mu$  or $\nu_\tau$)
neutrino flavors are found to have \emph{different} topologies with respect to $\bF$. 
This fact could have important implications for neutrino mixing [5].

With respect to the matrix transformation $\bF$,  the topology of both the 
$\nu_\mu$  and $\nu_\tau$
is found to be that of a M\"obius strip (Ref.\ 4, p.\ 143).  By contrast,  
the topology of the $\nu_e$
(with respect to $\bF$) is that of a cylinder.
And because we \emph{assume} that a change in topology 
during transitions tends to be suppressed,  the foregoing topological distinctions between 
neutrinos may help explain recent  observations of (nearly) 
maximal $\nu_\mu$-$\nu_\tau$  
mixing  [6, 7]. However, it is important to stress at the outset that there are good reasons for believing that similar topological distinctions among quarks do \emph{not} play an important role in $d$, $s$, $b$ quark mixing (See Ref.\ 8 and Sec.\ A.4.2). The principal purpose of 
this paper then, will be to focus attention almost exclusively on neutrinos, and to determine the constraints that the proposed topological distinctions 
place on \emph{conventional} $\nu_e, \nu_\mu, \nu_\tau$ \emph{neutrino} mixtures.

\section{Conventional Description of Neutrino Mixing}
At ``birth'' via weak decays, or upon detection via weak capture interactions, neutrinos have a \emph{definite} flavor and topology [9]. However, between birth and detection they are in a mixed state having no definite flavor or topology. In this intermediate region the probability of flavor (or topology) maintenance and/or change \emph{oscillates}. Only at great distances from the neutrino source do these oscillations finally ``damp out.''

In the conventional description of three-flavor neutrino mixing \emph{flavor eigenstates} are related to neutrino \emph{mass eigenstates} (states of definite mass) via a unitary 
Cabibbo-Kobayashi-Maskawa (CKM)-like ``mixing'' matrix $U_{\alpha i}$, as follows [10, 11]
\begin{equation}\label{eqn1}
\nu_\alpha=\sum^3_{i=1}U_{\alpha i}\nu_i.
\end{equation}
Here, $\nu_i=\nu_1, \nu_2$ and $\nu_3$ are mass eigenstates with mass eigenvalues $m_1, m_2$ and $m_3$, respectively, while $\nu_\alpha=\nu_e$, $\nu_\mu$ and $\nu_\tau$ are flavor eigenstates. Using a conventional parameterization
for $U_{\alpha i}$, (1) becomes [12]
\begin{equation}\label{eqn2}
\left(\begin{array}{c}
\nu_e\\
\nu_\mu\\
\nu_\tau\end{array}\right) =
\left(\begin{array}{ccc}
c_1 & s_1c_3, & s_1s_3\\
-s_1c_2,&c_1c_2c_3-s_2s_3e^{i\delta}, & c_1c_2s_3+s_2c_3e^{i\delta}\\
-s_1s_2, & c_1s_2c_3+c_2s_3e^{i\delta}, & c_1s_2s_3-c_2c_3e^{i\delta}\end{array}\right)\;
\left(\begin{array}{c}
\nu_1\\
\nu_2\\
\nu_3\end{array}\right),
\end{equation}
where $c_i\equiv\cos \theta_i$ and $s_i\equiv\sin\theta_i$, and $\delta$ is associated with a Dirac-type CP-noninvariant phase factor
$e^{i\delta}$.

Using (\ref{eqn2}) it can be shown that the probability of detecting a neutrino of flavor type $\beta$ at a distance $X$ from a source of neutrinos of flavor type $\alpha$ is given by [11]
\begin{equation}\label{eqn3}
P_{\nu_\alpha\to\nu_\beta} = \sum^3_{i=1}|U_{\alpha i}|^2|U_{\beta i}|^2 + \sum^3_{i\ne j} U_{\alpha i}U^\ast_{\beta i} 
U^\ast_{\alpha j}U_{\beta j}\cos\left(\frac{2\pi X}{l_{ij}}\right).
\end{equation}
Here, the so-called \emph{oscillation lengths} $l_{ij}$ are given by $l_{ij}=2\pi/(E_i-E_j)$, where the total relativistic
energy differences in a beam of neutrinos having \emph{fixed} momentum $p$ are $E_i-E_j=(m^2_i-m^2_j)/2p$.

Examination of (\ref{eqn3}) shows that at the neutrino source ($X=0$, and $t=0$) the probability $P_{\nu_\alpha\to\nu_\beta}$ reduces, as expected, to the $3\times 3$ identity matrix [13]
\begin{equation}\label{eqn4}
P_{\nu_\alpha\to\nu_\beta}\Bigg|_{X=0}=I_3,
\end{equation}
while at ``intermediate'' distances from the neutrino source $(X\approx l_{ij})$, the probability $P_{\nu_\alpha\to\nu_\beta}$ undergoes \emph{oscillations}.  Finally, at great distances from the neutrino source $(X\gg l_{ij})$, all time- or distance-dependent oscillations ``damp out,'' and we are left with a $3\times 3$ matrix of time-average probabilities [11], namely, 
\begin{equation}\label{eqn5}
\langle P_{\nu_\alpha\to\nu_\beta}\rangle = \sum^3_{i=1} |U_{\alpha i}|^2|U_{\beta i}|^2.
\end{equation}
From (\ref{eqn5}) it is clear that this matrix ($\alpha=$ row index, $\beta=$ column index), which describes long-distance neutrino mixtures, is \emph{symmetric}. Keeping in mind this symmetry, and calling this matrix $M$, one has
\begin{equation}\label{eqn6}
M = \left(\begin{array}{ccc}
\langle P_{\nu_e\to\nu_e}\rangle, & \langle P_{\nu_e\to\nu_\mu}\rangle, & \langle P_{\nu_e\to\nu_\tau}\rangle \\
\langle P_{\nu_\mu\to\nu_e}\rangle, & \langle P_{\nu_\mu\to\nu_\mu}\rangle, & \langle P_{\nu_\mu\to\nu_\tau}\rangle \\
\langle P_{\nu_\tau\to \nu_e}\rangle, & \langle P_{\nu_\tau\to\nu_\mu}\rangle, & \langle P_{\nu_\tau\to\nu_\tau}\rangle\end{array}\right).
\end{equation}
Note that all rows and columns of $M$ must sum to unity (total probability 1).

Using $M$ we can describe the expected neutrino flavor content at a great distance from a neutrino source (e.g., a supernova) as follows
\begin{equation}\label{eqn7}
\{D_e, D_\mu, D_\tau\}=M\{B_e, B_\mu, B_\tau\},
\end{equation}
where $\{\quad\}$ signifies column vectors, and $D_\alpha$ is the number of \emph{detected} neutrinos 
of definite flavor $\nu_\alpha$, and $B_\alpha$ their number at ``birth'' at some distant neutrino source. Note that because neutrinos are assumed to be conserved, the total number of neutrinos at birth equals their number upon ``detection,'' namely,
\begin{equation}\label{eqn8}
B_e+B_\mu+B_\tau=D_e+D_\mu+D_\tau.
\end{equation}

\section{Proposed Topological Constraints on Neutrino Mixing}
In this paper we will show, among other things, that certain proposed \emph{qualitative} topological constraints on the matrix of time-average probabilities $\lan P_{\nu_\alpha\to\nu_\beta}\ran$ or $M$, result in a \emph{quantitative} determination of certain  ($U$-matrix) neutrino mixing-parameters [14]. These mixing parameters, in turn, serve to provide a \emph{unique} determination of the matrix $\lan P_{\nu_\alpha\to\nu_\beta}\ran$ or $M$.

Given that the $\nu_e$ ($\nu_\mu$ or $\nu_\tau$) neutrino flavor
has the topology of a cylinder
(M\"obius strip) with respect to the internal transformation $\bF$,
and assuming that topological constraints are the
\emph{primary} determinants of the matrix $M$ describing long-distance neutrino 
mixtures, the \emph{form} of $M$ is easily determined.
Moreover, these same
topological constraints can be used to place numerical bounds on the
components of $M$. To accomplish these results we need only apply the
following very general principle to neutrino-neutrino transitions:
\medskip

\emph{All other things being equal, any neutrino flavor $\nu_\alpha$ {\rm(}i.e.,
$\nu_e$, $\nu_\mu$ or $\nu_\tau${\rm)}, which undergoes neutrino-neutrino
transitions that change neutrino topology, will tend to be suppressed,
while neutrino-neutrino transitions that maintain neutrino topology will
tend to be {\rm(}relatively{\rm)} enhanced} [5].
\medskip

To this principle we add the following corollary,
\medskip

\emph{All other things being equal, because the $\nu_\mu$ and $\nu_\tau$ neutrinos have
the same topology, they will act the same way in all neutrino-neutrino
transitions {\rm (}involving long-distance neutrino mixtures{\rm )}.}
\medskip

Given these principles,  and assuming as stated previously, that topological constraints are the \emph{primary} determinants of the matrix $M$, we immediately have the following  \emph{topological} constraints on
long-distance neutrino mixtures:
\medskip

A. No matter what neutrino flavor $(\nu_\alpha)$ and topology one starts with at
some distant source (say a supernova), by the time the neutrino 
mixture reaches its
``equilibrium'' state (where all time-dependent oscillations have ``damped out''), it should contain \emph{equal} fractions of $\nu_\mu$
and $\nu_\tau$, because these neutrinos have the \emph{same} topology.
\medskip

B. Because the $\nu_\mu$ and $\nu_\tau$ neutrinos have the \emph{same} topology, if one
starts out with \emph{either} a pure $\nu_\mu$ \emph{or} a pure $\nu_\tau$
source, one should end up with the \emph{same} long-distance equilibrium
mixture of $\nu_e$, $\nu_\mu$ and $\nu_\tau$.
\medskip

C. If topology is the controlling factor in describing long-distance neutrino mixtures, then \emph{there should be absolutely no {\rm(}effective{\rm)} difference between the two functions {\rm(}of $U$-matrix mixing parameters}), which describe $\lan P_{\nu_e\to\nu_\mu}\ran$ \emph{and} $\lan P_{\nu_e\to\nu_\tau}\ran$, because the $\nu_\mu$ and $\nu_\tau$ neutrinos have the \emph{same} topology.  Very loosely speaking we are assuming that these mathematical functions are
effectively ``topological invariants'' with respect to the exchange of flavor indices $\mu$ and $\tau$ (See Ref.\ 4, pp.\ 20 and 21). That is, not only are the two functions $\lan P_{\nu_e\to\nu_\mu}\ran$ and $\lan P_{\nu_e\to\nu_\tau}\ran$ required to be \emph{equal}, but they are also required to be equal, \emph{term-by-term} (i.e., they are required to be \emph{term-wise} equal). Similarly the three functions $\lan P_{\nu_\mu\to\nu_\mu}\ran$, $\lan P_{\nu_\mu\to\nu_\tau}\ran$ and $\lan P_{\nu_\tau\to\nu_\tau}\ran$ are required to be term-wise equal. 
\medskip

\setcounter{newsection}{3}

\subsection{The form of the matrix $M$}

Constraints A), B) and C) in the previous section, together with Eqs.\ (\ref{eqn7}) and (\ref{eqn8}), dictate that the symmetric
matrix $M$ describing long-distance neutrino mixtures must have the
form expressed by
\begin{equation}\label{eqn9}
\left(\begin{array}{c}
D_e\\
D_\mu\\
D_\tau\end{array}\right) = \left(
\begin{array}{ccc}
a&b&b\\
b&c&c\\
b&c&c\end{array}\right) 
\left(\begin{array}{c}
B_e\\
B_\mu\\
B_\tau\end{array}\right).
\end{equation}
Moreover,  the proposed topological constraints place numerical
bounds on the matrix elements $a$, $b$ and $c$. For example, because we are assuming that the $\nu_e$ is
\emph{inhibited}, but
 because of quantum fluctuations \emph{not} prevented, by its topology from turning into a $\nu_\mu$ or
$\nu_\tau$, we require the inequalities
\begin{equation}\label{eqn10}
a > b>0.
\end{equation}
Similarly, because we are assuming that  the $\nu_\mu$ and/or the $\nu_\tau$ are \emph{inhibited} by
their topology from turning into a $\nu_e$, we require the inequality
\begin{equation}\label{eqn11}
c > b.
\end{equation}

Equations (\ref{eqn10}) and (\ref{eqn11}), together with the requirement that
all rows and columns of the matrix $M$ in (\ref{eqn7}) and (\ref{eqn9}) sum to unity (i.e.,
$a+2b=b+2c=1$), further leads to the following inequalities
\begin{equation}\label{eqn12}
(\frac{1}{3}< a< 1),\qquad (0< b< \frac{1}{3}),\qquad (\frac{1}{3}<
c<\frac{1}{2}).
\end{equation}
\newpage

\noindent Moreover, since $a+2b=b+2c$, the arithmetic mean of $a$ and $b$ is $c$,
i.e., $(a+b)/2=c$, which means that $c$ lies half-way 
between $a$ and $b$. Hence,
the matrix elements $a$, $b$ and $c$ are subject to the combined inequalities
\begin{equation}\label{eqn13}
a> c> b>0.
\end{equation}
Equations (\ref{eqn9}), (\ref{eqn12}) and (\ref{eqn13}), together with items A), B) and C) above,  constitute the proposed topological constraints on the matrix $M$.
It happens that these constraints on $M$  are \emph{very} restrictive. In particular, we will show in the next section (See also Appendix B) that these constraints, together with the assumption that $\sin^2\delta\ne 1$ or $\cos^2\delta\ne 0$ (i.e., CP violation is \emph{not} maximal), admit of only \emph{one} solution for $M$!

\section{Determination of the Matrix $M$}
When the proposed topological constraints of Section 3.0 are applied to the conventional description of neutrino mixing (\ref{eqn2}), the matrix $M$ in (6), (7) and (9), is \emph{uniquely} determined. To see how this happens consider the following time-average probabilities (See Appendix B for details)
\begin{equation}\label{eqn14}
\langle P_{\nu_e\to\nu_\mu}\rangle = 2c^2_1 s^2_1 c^2_2 + 2s^2_1 s^2_3 c^2_3 (s^2_2 - c^2_1 c^2_2) + 2s^2_1 s_2s_3 c_1c_2c_3\cos \delta(s^2_3-c^2_3),
\end{equation}
and
\begin{equation}\label{eqn15}
\langle P_{\nu_e\to\nu_\tau}\rangle = 2 c^2_1 s^2_1 s^2_2 + 2s^2_1 s^2_3 c^2_3 (c^2_2 - c^2_1 s^2_2) - 2s^2_1 s_2 s_3 c_1 c_2 c_3 \cos\delta (s^2_3 - c^2_3).
\end{equation}
According to the proposed topological constraints of Section 3.0, these two  time-average probabilities must be 
\emph{term-wise equal}, and \emph{nonzero}. These requirements place \emph{three} constraints on the $U$-matrix mixing parameters, namely, $s^2_1>0$, $s^2_2=c^2_2=\frac{1}{2}$, and
\begin{equation}\label{eqn16}
s^2_1 s_2 s_3 c_1 c_2 c_3\cos\delta (s^2_3-c^2_3)=0.
\end{equation}
Note that (14) and (15) are term-wise equal as required, even with the minus sign preceeding the last term of (15) because this term is of zero magnitude.

Next, the topological constraint of Section 3.0, namely, $\langle P_{\nu_\mu\to\nu_\mu}\rangle = \langle P_{\nu_\tau\to\nu_\tau}\rangle$ can be realized provided the $U$-matrix
mixing parameters are further constrained by $s^2_3=c^2_3=\frac{1}{2}$, which also happens to satisfy (\ref{eqn16}). That is, given $s^2_2=c^2_2$ and $s^2_3=c^2_3$ one has (See Appendix B for details)
\begin{equation}\label{eqn17}
\langle P_{\nu_\mu\to\nu_\mu}\rangle = s^4_1 c^4_2 + 2c^4_2 c^4_3(c^2_1+1)^2 + 8c^2_1 c^4_2 c^4_3 \cos^2\delta,
\end{equation}
and
\begin{equation}\label{eqn18}
\langle P_{\nu_\tau\to\nu_\tau}\rangle = s^4_1 c^4_2 + 2c^4_2 c^4_3(c^2_1+1)^2 + 8c^2_1 c^4_2 c^4_3 \cos^2\delta.
\end{equation}

Now the topological constraints of Section 3.0 also require that $\langle P_{\nu_\mu\to\nu_\tau}\rangle$ be equal to (\ref{eqn17}) and (\ref{eqn18}). This places additional constraints on $M$, and the $U$-matrix mixing parameters.
Comparing (\ref{eqn17}) and (\ref{eqn18}) with the following expression from Appendix B for $\langle P_{\nu_\mu\to\nu_\tau}\rangle$, namely,
\begin{equation}\label{eqn19}
\langle P_{\nu_\mu\to\nu_\tau}\rangle = s^4_1 c^4_2 + 2c^4_2 c^4_3(c^2_1+1)^2 - 8c^2_1 c^4_2 c^4_3 \cos^2\delta,
\end{equation}
we see that the requisite equality $\langle P_{\nu_\mu\to\nu_\mu}\rangle = \langle P_{\nu_\tau\to\nu_\tau}\rangle = \langle P_{\nu_\mu\to\nu_\tau}\rangle$ leads to the constraint (note that $c^4_2=c^4_3>0$)
\begin{equation}\label{eqn20}
c^2_1 \cos^2\delta = 0.
\end{equation}
Assuming [15, 16] that $\sin^2\delta\ne 1$ or $\cos^2\delta\ne 0$ (i.e., CP violation is \emph{not} maximal),  
 we further determine from (20) that
\begin{equation}\label{eqn21}
c^2_1=0\mbox{ and }s^2_1=1.
\end{equation}
Note that (17), (18) and (19) exhibit the requisite term-wise equality [See item C) in Sec.\ 3.0] even with the minus sign preceeding the third term of (19), because this term is of zero magnitude.

Gathering together the predicted mixing-parameter constraints, namely, $c^2_2 = s^2_2$, $c^2_3 = s^2_3$, $s^2_1=1$ and $c^2_1=0$, we can express six of the time-average probabilities associated with the (symmetric) matrix $M$ as follows (Use Eqs. \ref{eqnb6}, \ref{eqn14}, \ref{eqn15} and \ref{eqn17}--\ref{eqn19})
\begin{equation}\label{eqn22}
\langle P_{\nu_e\to\nu_e}\rangle  = 1-2c^4_3, 
\end{equation}
\begin{equation}\label{eqn23}
\langle P_{\nu_e\to\nu_\mu}\rangle  =  \langle P_{\nu_e\to\nu_\tau}\rangle = 2c^2_2 c^4_3,  
\end{equation}
and
\begin{equation}\label{eqn24}
\langle P_{\nu_\mu\to\nu_\mu}\rangle  = \langle P_{\nu_\tau\to\nu_\tau}\rangle = \langle P_{\nu_\mu\to\nu_\tau}\rangle = 
c^4_2(1+2c^4_3). 
\end{equation}
But, the previous arguments have shown that the $U$-matrix mixing parameters satisfy $c^2_2=c^2_3 = \frac{1}{2}$, which leads to the following specific numerical \emph{predictions}
\begin{equation}\label{eqn25}
\langle P_{\nu_e\to\nu_e}\rangle  = \frac{1}{2} = a, 
\end{equation}
\begin{equation}\label{eqn26}
\langle P_{\nu_e\to\nu_\mu}\rangle  = \langle P_{\nu_e\to\nu_\tau}\rangle = \frac{1}{4} = b,  \end{equation}
and
\begin{equation}\label{eqn27}
\langle P_{\nu_\mu\to\nu_\mu}\rangle  = \langle P_{\nu_\tau\to\nu_\tau}\rangle = \langle P_{\nu_\mu\to\nu_\tau}\rangle = \frac{3}{8} = c. 
\end{equation}
Given that $M$ is symmetric, note that (\ref{eqn25}), (\ref{eqn26}) and (\ref{eqn27}) are \emph{the} only \emph{matrix elements, which could be consistent with the proposed topological constraints of Section 3.0, and the requirement that $\sin^2\delta\ne 1$ or  $\cos^2\delta\ne 0$} (See Ref.\ 15 and 16).

Employing Eqs. (\ref{eqn9}) and (\ref{eqn25}--\ref{eqn27}), one finally has the \emph{prediction} [6, 16]
\begin{equation}\label{eqn28}
M = \frac{1}{8} \left( \begin{array}{ccc}
4 & 2 & 2 \\
2 & 3 & 3 \\
2 & 3 & 3 \end{array}\right). 
\end{equation}
To summarize, \emph{the proposed} (\emph{qualitative}) \emph{topological constraints on the matrix $M$ {\rm (}See Sec.\ 3.0{\rm )} result in quantitative constraints on the $U$-matrix  mixing parameters {\rm (}See Eq. \ref{eqn2}{\rm )}, namely,} ($s^2_2=c^2_2=c^2_3 = s^2_3 = \frac{1}{2}, s^2_1=1$ \emph{or} $c^2_1=0$ \emph{with} $\sin^2\delta\ne 1$ or $\cos^2\delta\ne 0$ assumed), \emph{which in turn result in a} unique \emph{quantitative determination of the matrix} $M$ (See Eq. \ref{eqn28}).

\section{Topology-Maintaining and Topology-Changing Influences in Equilibrium?}

Not only is the matrix $M$ in (28) a \emph{unique} solution to the proposed topological constraints of Section 3.0 with $\sin^2\delta\ne 1$ or $\cos^2\delta\ne 0$, but it also has some very special properties that may eventually help reveal the deeper
\emph{dynamical} significance of this matrix. Note that (6), (9), and (28) have the very special property 
\begin{equation}\label{eqn501}
\langle P_{\nu_e\to\nu_e}\rangle = \langle P_{\nu_e\to\nu_\mu}\rangle + \langle P_{\nu_e\to\nu_\tau}\rangle.
\end{equation}
This equation says that the time-average probability that the $\nu_e$ topology  \emph{doesn't} change, namely,
\begin{equation}\label{eqn502}
P_{NC}=\langle P_{\nu_e\to\nu_e}\rangle = a, 
\end{equation}
and the time-average probability that the $\nu_e$ topology \emph{does} change, namely,
\begin{equation}\label{eqn503}
P_C=\langle P_{\nu_e\to\nu_\mu}\rangle + \langle P_{\nu_e\to\nu_\tau}\rangle = (1-a), 
\end{equation}
are \emph{equal}, namely,
\begin{equation}\label{eqn504}
P_C=P_{NC}.
\end{equation}

Now this equality looks very much like an ``equilibrium'' condition between those underlying physical influences that would act to \emph{change} the $\nu_e$ topology (quantum fluctuations), and those underlying physical influences that would act to \emph{maintain} the $\nu_e$ topology (e.g., energy ``barriers'').

We will now provide further support for this proposal. In particular, consider the ``joint'' probability [17]
\begin{equation}\label{eqn505}
P=P_C\cdot P_{NC},
\end{equation}
and notice that
\begin{equation}\label{eqn506}
\frac{d P}{d a} = P_C\;\frac{d P_{NC}}{da} + P_{NC}\;\frac{dP_C}{d a}.
\end{equation}
From (30) and (31) this last equation reduces to
\begin{equation}\label{eqn507}
\frac{dP}{da} = P_C-P_{NC}.
\end{equation}
And, taking the second derivative, we also find
\begin{equation}\label{eqn508}
\frac{d^2P}{d a^2} = -2<0.
\end{equation}

Therefore, when $P_C=P_{NC}$ we discover that the joint probability $P=P_C\cdot P_{NC}$ is a \emph{maximum}, namely, it characterizes some \emph{most probable} condition or ``state.'' And, this of course is the very essence of an ``equilibrium'' condition [7].
However, it must be understood that this hypothetical (long-distance) ``equilibrium'' between topology-changing, and topology-maintaining physical influences, is only a (cumulative) \emph{result} of deeper, and largely unknown (short-distance), dynamical  processes in the vacuum, which first begin to
\emph{act on neutrinos at their source---thereby, eventually establishing the equilibrium condition $P_C=P_{NC}$---on time scales very much shorter than the time it takes for the time-dependent oscillations in neutrino mixtures to ``damp out.''}  This is an essential requirement if these 
hypothetical short-distance processes are to be responsible for ``selecting'' the (constant) $U$-matrix mixing parameters \emph{prior} to neutrino mixing.

Finally, while the \emph{unique} nature of (28) has been demonstrated, and while we see that this result seems  to be closely connected with (short-distance) physical processes involving \emph{topological constraints} that \emph{select} the neutrino mixing parameters, it is also clear that we do not possess a first-principles formulation of these ideas. As such, they are only promising conjectures.

\section{Conclusions}

Topological constraints could play a major role in determining the 
nature of  neutrino mixtures.  Given that the
$\nu_e$ and ($\nu_\mu$ or $\nu_\tau$) neutrino flavors have \emph{distinct} topologies,
and assuming that 
topology changes are suppressed in neutrino-neutrino 
transitions, while neutrino-neutrino 
transitions without topology-change are relatively enhanced---one easily 
determines both the form of the matrix describing
long-distance neutrino mixtures, and \emph{uniquely} determines the numerical magnitudes of the individual matrix elements.
If the predicted matrix (28),
is eventually verified by observations 
of neutrinos from distant astronomical sources (e.g., supernovae), and/or in long-baseline terrestrial experiments,
this will provide qualitative
support for the new description of fundamental fermions, which 
requires, among other things, that
the $\nu_e$ and ($\nu_\mu$  or $\nu_\tau$) neutrino flavors start, and end, ``life''
as \emph{topologically distinct} quantum objects (See Appendix A and Refs.\ 3--5). However, while experiments could certainly falsify (28), (9), (12) and (13), together with items A), B) and C) in Section 3.0, and the assumption [15, 16] that CP violation is \emph{not} maximal (i.e., $\sin^2\delta\ne 1$ or $\cos^2\delta\ne 0$), experimental verification of these things would \emph{not} confirm the proposed \emph{topological} constraints of Section 3.0. A much better theoretical understanding of the dynamical significance (if any) of the new 2-space description of quarks and leptons will be required before one can be confident that, if verified, (28) really is a reflection of underlying topological constraints [18].

\section{Acknowledgments}

I am indebted to F.\ Wietfeldt and R.\ Zannelli for numerous helpful discussions. I also thank M.\ Sheetz for manuscript preparation and submission.


\newpage

\renewcommand \thesection{\arabic{section}.0}
\renewcommand \thesubsection{8.\arabic{subsection}}
\renewcommand \thesubsubsection{A.\arabic{subsubsection}}

\setcounter{section}{7}

\subsection*{Appendix A. A new Internal 2-Space Description of Quarks and Leptons}

\setcounter{newnewsection}{0}
\renewcommand{\theequation}{A\arabic{equation}}
\setcounter{equation}{0}

\subsubsection{Background}

In spite of the many successes of the standard model of particle physics,
the
observed proliferation of matter-fields, in the form of ``replicated"
generations or families, is a major unsolved problem.  In [3, 19 and 20]
a new organizing principle for fundamental fermions was proposed, i.e.,
a minimalistic
``extension" of the standard model based, in part, on the Cayley-Hamilton
theorem for matrices.   To introduce (internal) global degrees
of freedom that are capable of distinguishing all observed flavors, the
Cayley-Hamilton theorem was used
to generalize the familiar standard-model concept of
scalar fermion-numbers $f$
 (i.e., $f_m=+1$ for all fermions and $f_a=-1$ for
all
antifermions).  This theorem states that every (square) matrix satisfies its
characteristic equation. Hence, if $f_m$ and $f_a$
are taken to be the eigenvalues
of some real matrix $\bF$ (a ``generalized fermion  number"), it follows
from
this theorem that both $f$ and $\bF$ are square-roots of unity.  Assuming
further that the components of both $\bF$ and its eigenvectors are global
charge-like quantum observables, and that $\bF$ ``acts" on a (real) internal vector
2-space, both the form of $\bF$
and the 2-space metric are determined.  One finds that the 2-space has a
non-Euclidean or ``Lorentzian" metric, and that various associated 2-scalars
serve as global flavor-defining ``charges," which can be identified with
Lorentz 4-scalar charges such as strangeness, charm, baryon and lepton numbers.  

Because these global charges are essentially the global charges associated with the so-called ``accidental symmetries'' of the Lagrangian describing strong and electroweak interactions [19--21], they
 can be used to describe individual flavors (i.e.,
flavor
eigenstates), flavor doublets and families.  Moreover, because of the
aforementioned non-Euclidean constraints, and certain standard-model
constraints, one finds
that these global charges are effectively ``quantized" in
such a way that families are replicated.  Finally, because these same
constraints dictate that there are only a limited number of values these
charges can assume, one finds that families, and their associated neutrinos,
always come in ``threes."

\subsubsection{Representing Flavor Eigenstates}

 The eigenvectors $\bQ$ of $\bF$ (i.e., $\bF\bQ=f\bQ$ where $f$ is the
scalar fermion-number), together with certain pairs of linearly independent
vectors ($\bU$ and $\bV$) that resolve $\bQ$ (i.e., $\bQ=\bU+\bV$), namely,
various non-Euclidean vector ``triads'' $(\bQ, \bU, \bV)$---these are the
analogs of Euclidean triangles---serve to represent flavor-doublets in terms
of a pair of quark or lepton flavor-eigenstates as follows:
\begin{equation}
|\hbox{``up''}\rangle = |q_1, u_1, v_1, {\mathbf{Q}}^2, {\mathbf{U}}^2,
2\bU\bullet\bV\rangle
\end{equation}
and
\begin{equation}
|\hbox{``down''}\rangle = |q_2, u_2, v_2, {\mathbf{Q}}^2,
{\mathbf{U}}^2, 2\bU\bullet\bV\rangle.
\end{equation}
Here, $\bQ=\{q_1, q_2\}$, $\bU=\{u_1, u_2\}$ and $\bV=\{v_1,v_2\}$ are column-%
vectors and their components $q_1, q_2, u_1, u_2$, $v_1$ and $v_2$,
together with the non-Euclidean scalar products $\bQ^2$, $\bU^2$, $\bV^2$ and $\bU\bullet\bV$, 
 are various global mutually-commuting flavor-defining charge-like
quantum numbers (e.g., $q_1$ and $q_2$ are electric charges carried by ``up''- and ``down''-type flavors in a flavor doublet).

\subsubsection{Topology of Vector Triads}

When we refer to the ``topology'' of a particular neutrino flavor-eigenstate
(e.g., the $\nu_e$), we are referring to the topology of the corresponding
\emph{vector triad} $(\bQ, \bU, \bV)$ with respect to the internal
transformation $\bF$. And, because $\bF$ generates the M\"obius group $Z_2$
(i.e., $\bF^2=\bI_2)$, those vector triads that are left \emph{unchanged} by
$\bF$, and by the identity $\bF^2=\bI_2$, [i.e., symbolically $\bF(v)(\bQ, \bU, \bV)\equiv (\bQ, \bU, \bV)$],  have the (abstract) topology of a \emph{cylinder}, while vector triads that are
\emph{changed} by $\bF$ [i.e., $\bF(v)(\bQ, \bU, \bV)\not\equiv (\bQ, \bU, \bV)$], but obviously not changed by the identity $\bF^2=\bI_2$, have the (abstract)
topology of a \emph{M\"obius strip}. And, as it turns out, the neutrino
flavor $\nu_e$ ($\nu_\mu$ or $\nu_\tau$) corresponds to a vector triad 
having the topology of a cylinder (M\"obius strip) with respect to $\bF$. See the qualifying remarks in the next section.

\subsubsection{Qualifying Remarks}

With the possible exception of one family, \emph{all quarks and leptons
within a given family are found to exhibit the same topology with respect
to} $\bF$. This is certainly the case for the \emph{first} and \emph{third}
families. And, because of these facts, together with the requirement of
quark-lepton ``universality,'' at least within any given family, we
naturally expect this to be the case for the \emph{second} family as well.
It happens that the second-family $c$ and $s$ quarks exhibit M\"obius
topology with respect to $\bF$. However, strictly speaking, the second-family
leptons, namely, the muon and its associated neutrino (and associated
antiparticles) exhibit the requisite M\"obius topology \emph{only} if the
components of the associated $\bV$-vectors are \emph{not} exactly zero (See
p.\ 57, and Tables II and IV in Ref.\ 3 where the muon $\bV$-vectors, which are functions of the strong-color multiplicity $M_c$,  are shown to be $\bV\equiv 0$ in the limit where $M_c\equiv 1$). Therefore, if we wish to maintain a kind of quark-lepton
``universality'' within families, the $\bV$-vector components associated
with muons, though  very small, must be \emph{nonzero}.  And, according to the 2-space mathematical description of $\bV$-vectors, \emph{this is always the case whenever} $M_c\, \lower3pt\hbox{$\buildrel >\over\sim$}\, 1$.

\subsubsection*{A.4.1\quad$\nu_\mu$ neutrinos with $M_c\, \lower3pt\hbox{$\buildrel >\over\sim$}\, 1$.}

It has been shown (See Ref.\ 3, pp.\ 52--55) that the quark and lepton electric charges are the 
``up''-``down'' components of the \emph{eigenvectors} of the matrix $\bF(v)$, where $v=\ln M_c$. In particular, the quark charges are given by $(M_c=3)$
\begin{eqnarray}
q_1(f) & = & \frac{(M^2_c-1)}{2M_c(M_c-f)} = +\frac{2}{3}\hbox{ for }f=+1\hbox{ and }+\frac{1}{3}\hbox{ for }f=-1, \label{eqn55} \\
q_2(f) & = & q_1(f)-1, \label{eqn56}
\end{eqnarray}
where the baryon number for quarks is $B=q^2_1(f)-q^2_2(f)=\pm\frac{1}{3}$ for $f=\pm 1$.
Similarly, the lepton electric charges are given by $(M_c=1$)
\begin{eqnarray}
q'_1(f) & = & \frac{-(M_c^2-1)}{2M_c(M_c-f)} = -1\hbox{ for }f=+1\hbox{ and }0\hbox{ for }f=-1, \label{eqn57} \\
q'_2(f) & = & q'_1(f)+1, \label{58}
\end{eqnarray}
where the lepton number for leptons is $L=[q'_1(f)]^2-[q'_2(f)]^2=\pm 1$ for $f=\pm 1$.

Now, if the eigenvector $\bQ$ of $\bF(v)$ is $\bQ=\{q'_1(f), q'_2(f)\}$, and $f=+1$ for leptons where $\bQ=\bU+\bV$, we have for leptons
\begin{equation}
\bV=\bQ-\bU,
\end{equation}
where the 2-vectors $\bU, \bV$ are appropriate to a description of leptons. According to the 2-space description, $\bQ$ and $\bV$ are \emph{both} functions of $M_c$, while $\bU$ is not. Only in the case of the muons ($\mu^-$ and $\nu_\mu$) with $M_c\equiv 1$ is $\bQ\equiv \bU\equiv \{-1,0\}$ and $\bV\equiv 0$. In this limiting case, the vector triad $(\bQ, \bU, \bV)$ does \emph{not} possess the requisite second-family M\"obius topology. In fact, it transforms under $\bF$ like a \emph{cylinder}. So, a way to insure that $\bV\ne 0$, which lies completely within the context of the existing 2-space mathematical description, is to treat $M_c$ as a continuous \emph{variable}, and let $M_c$ closely approach, but never actually equal one in (A5),  (A6), and (A7) while maintaining $\bU=\{-1,0\}$. In this case,  $\bV\ne 0$, which leads to the requisite second-family 
M\"obius topology for the vector triad $(\bQ, \bU, \bV)$, and hence the $\nu_\mu$ neutrino.
Of course, when $M_c\, \lower3pt\hbox{$\buildrel >\over\sim$}\, 1$, neutrinos, including the $\nu_\mu$, would carry an \emph{infinitesimal} electric charge $q'_2(f)\, \lower3pt\hbox{$\buildrel >\over\sim$}\, 0$.  However, from a ``practical'' physical standpoint this is not a serious problem because $q'_2(f)$  \emph{can always be made arbitrarily small, while always maintaining the requisite second-family M\"obius topology of the vector triad associated with the $\nu_\mu$ neutrino}.

It is clear from the foregoing discussion that the current \emph{mathematical} and \emph{physical} description of the 2-space is simply \emph{not} ``robust'' enough to demonstrate, without a doubt, that $\bV\ne 0$ for the $\nu_\mu$ neutrino. Nevertheless, for various reasons outlined at the beginning of this subsection (A.4), and because of the foregoing mathematical justification, we will assume throughout this paper that this is indeed the case, and that, as a consequence, the vector triad associated with the $\nu_\mu$ neutrino does possess the requisite second-family M\"obius topology with respect to $\bF$.

\subsubsection*{A.4.2\quad Topological influences in quark mixing.}

As described above, all quarks and leptons within a given family exhibit the \emph{same} topology with respect to the internal transformation $\bF$. However, as described in [8], owing to the uncertainty principle, and the relatively large mass differences between quarks, mixing among the $d$, $s$, $b$ quarks (inside quark composites) is \emph{not} expected to be controlled by topology. By contrast, $\nu_e, \nu_\mu, \nu_\tau$ neutrino mixing \emph{is} expected to be controlled by topology, owing to the small value, and near degeneracy, of neutrino masses.

\newpage

\renewcommand \thesection{\arabic{section}.0}
\renewcommand \thesubsection{8.\arabic{subsection}}
\renewcommand \thesubsubsection{B.\arabic{subsubsection}}

\setcounter{section}{8}

\subsection*{Appendix B. Determination of $\langle P_{\nu_\alpha\to\nu_\beta}\rangle$.}

\setcounter{section}{2}

\renewcommand{\theequation}{\thenewnewsection\arabic{equation}}
\setcounter{equation}{0}

The standard description of three-flavor neutrino mixtures [10, 11] involves a CKM-like unitary matrix $U_{\alpha i}$  that transforms \emph{mass eigenstates} $\nu_i=(\nu_1, \nu_2, \nu_3)$ into \emph{flavor eigenstates} $(\nu_e, \nu_\mu, \nu_\tau)$ as follows (See Eqs. \ref{eqn1} and \ref{eqn2} in the main text)
\begin{equation}\label{eqnb1}
\nu_\alpha = \sum^3_{i=1} U_{\alpha i}\nu_i.
\end{equation}
Using this formulation one can calculate the time-dependent probability $P_{\nu_\alpha\to\nu_\beta}$ that a neutrino of some definite flavor $\alpha$ (at ``birth'') will transform into a neutrino of a generally different flavor $\beta$, upon detection. The result is (\ref{eqn3}) in the main text.

At sufficiently great distances from the source of neutrinos, the time-dependent part of (\ref{eqn3}) ``damps out,'' and we are left with the time-independent or time-average probability
\begin{equation}\label{eqnb2}
\langle P_{\nu_\alpha\to\nu_\beta}\rangle = \sum^3_{i=1}|U_{\alpha i}|^2|U_{\beta i}|^2.
\end{equation}
Using (\ref{eqnb2}), and the $U$-matrix given by (\ref{eqn2}) in the main text, we can calculate all $\langle P_{\nu_\alpha\to\nu_\beta}\rangle$. For example, we find (Note that $e$, $\mu$ and $\tau$ refer to $U$-matrix \emph{row} indices 1, 2 and 3, respectively)
\begin{eqnarray}
\langle P_{\nu_e\to\nu_e}\rangle & = & \sum^3_{i=1} |U_{1i}|^4 = |U_{11}|^4 + |U_{12}|^4 + |U_{13}|^4 \hspace{2in}\\[3mm] \label{eqnb3}
& = & c^4_1 + s^4_1 c^4_3 + s^4_1 s^4_3 \\[3mm] \label{eqnb4}
& = & c^4_1 + s^4_1(1-2c^2_3 s^2_3), \\[3mm] \label{eqnb5}
\langle P_{\nu_e\to\nu_e}\rangle & = & 1-2c^2_1 s^2_1 - 2s^4_1 s^2_3 c^2_3. \label{eqnb6}
\end{eqnarray}
\noindent Similarly, we have
\begin{eqnarray}
\langle P_{\nu_e\to\nu_\mu}\rangle & = & \sum^3_{i=1} |U_{1i}|^2|U_{2i}|^2 = |U_{11}|^2|U_{21}|^2 + |U_{12}|^2|U_{22}|^2 + |U_{13}|^2|U_{23}|^2, \\ \label{eqnb7}
& = & c^2_1 s^2_1 c^2_2 + s^2_1 c^2_3 |c_1c_2c_3-s_2s_3\cos\delta-i s_2s_3\sin\delta|^2 \nonumber \\
&&\qquad + \; s^2_1s^2_3|c_1c_2s_3 + s_2c_3 \cos\delta + i s_2c_3 \sin\delta|^2,  \\ \label{eqnb8}
& = & s^2_1\{c^2_1c^2_2 + c^2_3[(c_1c_2c_3-s_2s_3\cos\delta)^2 + s^2_2 s^2_3 \sin^2\delta] \nonumber \\
&& \qquad +\; s^2_3[(c_1c_2s_3+s_2c_3\cos\delta)^2 + s^2_2c^2_3\sin^2\delta]\},\\ \label{eqnb9}
& = & s^2_1\{c^2_1c^2_2 + c^2_3[c^2_1c^2_2c^2_3 - 2c_1c_2c_3s_2s_3\cos\delta  +s^2_2s^2_3\cos^2
\delta + s^2_2 s^2_3\sin^2\delta] \nonumber \\
&& \qquad +\; s^2_3[c^2_1c^2_2s^2_3+2c_1c_2s_3s_2c_3\cos\delta + s^2_2c^2_3\cos^2\delta + s^2_2c^2_3\sin^2\delta]\}, \\ \label{eqnb10}
& = & s^2_1 \{ c^2_1c^2_2 + c^2_1c^2_2(c^4_3+s^4_3) - 2c_1c_2c_3s_2s_3\cos\delta (c^2_3) \nonumber \\
&& \qquad +\; 2c_1c_2c_3s_2s_3\cos\delta(s^2_3) + 2c^2_3s^2_2s^2_3\}, \\ \label{eqnb11}
& = & s^2_1\{c^2_1c^2_2 + c^2_1c^2_2(1-2c^2_3s^2_2)+2c^2_3s^2_2s^2_3 \nonumber \\
&& \qquad +\;2c_1c_2c_3s_2s_3\cos\delta (s^2_3-c^2_3)\}, \\ \label{eqnb13}
& = & s^2_1\{2c^2_1c^2_2 + 2c^2_3s^2_2s^2_3 - 2c^2_1c^2_2c^2_ss^2_3 \nonumber \\
&& \qquad +\; 2c_1c_2c_3s_2s_3\cos\delta(s^2_3-c^2_3)\}, \\ \label{eqnb12}
& = & s^2_1\{2c^2_1c^2_2+2c^2_3s^2_3(s^2_2-c^2_1c^2_2) + 2c_1c_2c_3s_2s_3\cos\delta(s^2_3-c^2_3)\}, \\ \label{eqnb14}
\langle P_{\nu_e\to\nu_\mu}\rangle & = & 2c^2_1s^2_1c^2_2 + 2s^2_1s^2_3c^2_3(s^2_2- c^2_1c^2_2) + 2s^2_1s_2s_3c_1c_2c_3\cos\delta(s^2_3-c^2_3). \label{eqnb15}
\end{eqnarray}
And,
\begin{eqnarray}
\langle P_{\nu_e\to\nu_\tau}\rangle & = & \sum^3_{i=1} |U_{1i}|^2|U_{3i}|^2 = |U_{11}|^2|U_{31}|^2 + |U_{12}|^2|U_{32}|^2 + |U_{13}|^2|U_{33}|^2, \\ \label{eqnb16}
& = & c^2_1 s^2_1 s^2_2 + s^2_1 c^2_3 |c_1s_2c_3+c_2s_3\cos\delta +ic_2s_3\sin\delta|^2 \nonumber \\
&& \qquad + s^2_1s^2_3|c_1s_2s_3 - c_2c_3\cos\delta -ic_2c_3\sin\delta|^2, \\ \label{eqnb17}
& = & s^2_1\{ c^2_2s^2_2 + c^2_3[(c_1s_2c_3+c_2s_3\cos\delta)^2 + c^2_2 s^2_3\sin^2\delta] \nonumber \\
&&\qquad + s^2_3[(c_1s_2s_3-c_2c_3\cos\delta)^2 + c^2_2c^2_3\sin^2\delta]\}, \\ \label{eqnb18}
& = & s^2_1\{c^2_1s^2_2 + c^2_3[c^2_1s^2_2c^2_3 + 2c_1c_2c_3s_2s_3\cos\delta + c^2_2 s^2_3\cos^2\delta + c^2_2s^2_3\sin^2\delta] \nonumber \\
&&\qquad + s^2_3[c^2_1s^2_2s^2_3-2c_1c_2c_3s_2s_3\cos\delta + c^2_2c^2_3\cos^2\delta + c^2_2c^2_3\sin^2\delta]\}, \\ \label{eqnb19}
& = & s^2_1 
\{ c^2_1s^2_2 + c^2_1 s^2_2 (c^4_3+s^4_3) + 2c_1c_2c_3s_2s_3\cos\delta(c^2_3) \nonumber \\
&& \qquad - 2c_1c_3c_3s_2s_3\cos\delta(s^2_3) + 2c^2_3 c^2_2 s^2_3\}, \\ \label{eqnb20}
& = & s^2_1 \{c^2_1s^2_2 + c^2_1s^2_2(1-2c^2_3s^2_3) + 2c^2_3c^2_2s^2_3 - 2c_1c_2c_3s_2s_3\cos\delta (s^2_3-c^2_3)\}, \\ \label{eqnb21}
& = & s^2_1\{2c^2_1s^2_2+2c^2_3c^2_2s^2_3 - 2c^2_1s^2_2c^2_3s^2_3 - 2c_1c_2c_3s_2s_3\cos\delta (s^2_3-c^2_3)\}, \\ \label{eqnb22}
& = & s^2_1 \{2c^2_1s^2_2 + 2c^2_3s^2_3(c^2_2-c^2_1s^2_2) - 2c_1c_2c_3s_2s_3\cos\delta (s^2_3-c^2_3) \}, \\ \label{eqnb23}
\langle P_{\nu_e\to\nu_\tau}\rangle & = & 2c^2_1s^2_1s^2_2+2s^2_1s^2_3c^2_3(c^2_2-c^2_1s^2_2)-2s^2_1 s_2s_3c_1c_2c_3\cos\delta(s^2_3-c^2_3). \label{eqnb24}
\end{eqnarray}
Now, according to the proposed topological constraints of Section 3.0 in the main text, (B15) and (\ref{eqnb24}) must be \emph{term-wise equal} and \emph{nonzero}. These conditions yield \emph{three} constraints on the mixing parameters, namely,
\begin{gather}
s^2_1  >  0, \\ 
s^2_2 =  c^2_2 = \frac{1}{2}, 
\end{gather}
and
\begin{equation}\label{eqnb27}
s^2_1s_2s_3c_1c_2c_3\cos\delta(s^2_3-c^2_3)=0.
\end{equation}

Next, according to the proposed topological constraints of Section 3.0 in the main text, the matrix elements $\langle P_{\nu_\mu\to\nu_\mu}\rangle$ and $\langle P_{\nu_\tau\to\nu_\tau}\rangle$ must be \emph{equal}. As we will now demonstrate, this condition yields another constraint on the mixing parameters, namely,
\begin{equation}\label{eqnb28}
s^2_3=c^2_3=\frac{1}{2},
\end{equation}
which also happens to satisfy (\ref{eqnb27}).
\begin{eqnarray}
\langle P_{\nu_\mu\to\nu_\mu}\rangle & = & \sum^3_{i=1}|U_{2i}|^4 =|U_{21}|^4+ |U_{22}|^4 + |U_{23}|^4, \\ \label{eqnb29}
& = & s^4_1c^4_2 + [(c_1c_2c_3-s_2s_3\cos\delta)^2 + s^2_2 s^2_3\sin^2\delta]^2 \nonumber \\
&&\qquad + [(c_1c_2s_3+s_2c_3\cos\delta)^2 + s^2_2c^2_3\sin^2\delta]^2, \\ \label{eqnb30}
& = & s^4_1c^4_2 + [c^2_1c^2_2c^2_3-2c_1c_2c_3s_2s_3\cos\delta + s^2_2s^2_3\cos^2\delta + s^2_2s^2_3\sin^2\delta]^2 \nonumber \\
&&\qquad + [c^2_1c^2_2s^2_3+2c_1c_2s_3s_2c_3\cos\delta + s^2_2c^2_3\cos^2\delta + s^2_2c^2_3\sin^2\delta]^2. \label{eqnb31}
\end{eqnarray}
Using $s^2_2=c^2_2$ and $c^2_3=s^2_3$
\begin{eqnarray}
\langle P_{\nu_\mu\to\nu_\mu}\rangle & = & s^4_1c^4_2 + [c^2_1c^2_2c^2_3-2c_1c_2c_3s_2s_3\cos\delta + c^2_2c^2_3\cos^2\delta + c^2_2 c^2_3 \sin^2\delta]^2 \nonumber \\
&&\qquad + [c^2_1c^2_2c^2_3+2c_1c_2s_3s_2c_3\cos\delta + c^2_2c^2_3\cos^2\delta + c^2_2c^2_3\sin^2\delta]^2, \\ \label{eqnb32}
& = & s^4_1c^4_2 + [(c^2_1c^2_2c^2_3+c^2_2c^2_3) - 2c_1c_2c_3s_2s_3\cos\delta]^2 \nonumber \\
&& \qquad + [(c^2_1c^2_2c^2_3+c^2_2c^2_3) + 2c_1c_2c_3s_2s_3\cos\delta]^2, \\ \label{eqnb33}
& = & s^4_1c^4_2+2(c^2_1c^2_2c^2_3+c^2_2c^2_3)^2 + 2(2c_1c_2c_3s_2s_3\cos\delta)^2, \\ \label{eqnb34}
& = & s^4_1c^4_2 + 2c^4_2c^4_3(c^2_1+1)^2 + 8c^2_1c^2_2c^2_3s^2_2s^2_3\cos^2\delta, \\ \label{eqnb35}
\langle P_{\nu_\mu\to\nu_\mu}\rangle & = & s^4_1c^4_2 + 2c^4_2c^4_3(c^2_1+1)^2+8c^2_1c^4_2c^4_3\cos^2\delta. \label{eqnb36}
\end{eqnarray}
Similarly, with a few steps removed, and once again using $s^2_2=c^2_2$, $c^2_3=s^2_3$ 
\begin{eqnarray}
\langle P_{\nu_\tau\to\nu_\tau}\rangle & = & \sum^3_{i=1} |U_{3i}|^4 = |U_{31}|^4 + |U_{32}|^4 + |U_{33}|^4, \\ \label{eqnb37}
& = & s^4_1s^4_2 + [(c_1s_2c_3+c_2s_3\cos\delta)^2 + (c_2s_3\sin\delta)^2]^2 \nonumber \\
&&\qquad + [(c_1s_2s_3-c_2c_3\cos\delta)^2+(c_2c_3\sin\delta)^2]^2, \\ \label{eqnb38}
&= & s^4_1c^4_2 + [c^2_1c^2_2c^2_3+2c_1c^2_2c^2_3\cos\delta + c^2_2c^2_3\cos^2\delta + c^2_2c^2_3\sin^2\delta]^2 \nonumber \\
&&\qquad + [c^2_1c^2_2c^2_3-2c_1c^2_2c^2_3\cos\delta+c^2_2c^2_3\cos^2\delta+c^2_2c^2_3\sin^2\delta]^2,\hspace{.3in} \\ \label{eqnb39}
& = & s^4_1c^4_2 + [(c^2_1c^2_2c^2_3+c^2_2c^2_3) + (2c_1c^2_2c^2_3\cos\delta)]^2 \nonumber \\
&&\qquad + [c^2_1c^2_2c^2_3+c^2_2c^2_3 - 2c_1c^2_2c^2_3\cos\delta]^2, \\ \label{eqnb40}
& = & s^4_1c^4_2 + 2(c^2_1c^2_2c^2_3+c^2_2c^2_3)^2 + 2(2c_1c^2_2c^2_3\cos\delta)^2, \\ \label{eqnb41}
\langle P_{\nu_\tau\to\nu_\tau}\rangle & = & s^4_1c^4_2 + 2c^4_2c^4_3(c^2_1+1)^2 + 8c^2_1c^4_2c^4_3\cos^2\delta. \label{eqnb42}
\end{eqnarray} 

Clearly, (\ref{eqnb41}) and (\ref{eqnb36}) are properly term-wise equal, as required by the topological constraints of Section 3.0. 

Thus far, we have established the following constraints on the mixing parameters ($s^2_2=c^2_2$, $s^2_3=c^2_3$ and $s^2_1>0$). Using these constraints we will now calculate
$\langle P_{\nu_\mu\to\nu_\tau}\rangle$.
\begin{eqnarray}
\langle P_{\nu_\mu\to\nu_\tau}\rangle & = & \sum^3_{i=1}|U_{2i}|^2|U_{3i}|^2 = |U_{21}|^2|U_{31}|^2 + |U_{22}|^2|U_{32}|^2 + |U_{23}|^2|U_{33}|^2, \\ \label{eqnb43}
& = & s^4_1c^4_2 + |c_1c_2c_3-s_2s_3e^{i\delta}|^2|c_1s_2c_3+c_2s_3e^{i\delta}|^2 \nonumber \\
&&\qquad + |c_1c_2s_3+s_2c_3e^{i\delta}|^2|c_1s_2s_3-c_2c_3e^{i\delta}|^2, \\ \label{eqnb44}
& = & s^4_1c^4_2+|c_1c_2c_3-s_2s_3\cos\delta-is_2s_3\sin\delta|^2|c_1s_2c_3+c_2s_3\cos\delta + ic_2s_3\sin\delta|^2 \nonumber \\
&&\qquad +|c_1c_2s_3+s_2c_3\cos\delta + is_2c_3\sin\delta|^2 |c_1s_2s_3-c_2c_3\cos\delta-ic_2c_3\sin\delta|^2,\hspace{.35in} \\ \label{eqnb45}
& = & s^4_1c^4_2 + [c^2_1c^2_2c^2_3-2c_1c^2_2c^2_3\cos\delta + c^2_2c^2_3\cos^2\delta + c^2_2c^2_3\sin^2\delta] \nonumber \\
&&\qquad\times [c^2_1c^2_2c^2_3+2c_1c^2_2c^2_3\cos\delta + c^2_2c^2_3\cos^2\delta + c^2_2c^2_3\sin^2\delta] \nonumber \\
&&\qquad + [c^2_1c^2_2c^2_3+2c_1c^2_2c^2_3\cos\delta + c^2_2c^2_3\cos^2\delta + c^2_2c^2_3\sin^2\delta] \nonumber \\
&&\qquad \times [c^2_1c^2_2c^2_3-2c_1c^2_2c^2_3\cos\delta + c^2_2c^2_3\cos^2\delta + c^2_2c^2_3\sin^2\delta], \\ \label{eqnb46}
& = & s^4_1c^4_2 + [c^2_1c^2_2c^2_3 + c^2_2c^2_3 - 2c_1c^2_2c^2_3\cos\delta]  [c^2_1c^2_2c^2_3+c^2_2c^2_3+2c_1c^2_2c^2_3\cos\delta] \nonumber \\
&&\qquad + [c^2_1c^2_2c^2_3+c^2_2c^2_3 + 2c_1c^2_2c^2_3\cos\delta]  [c^2_1c^2_2c^2_3 + c^2_2c^2_3 - 2c_1c^2_2c^2_3\cos\delta], \\ \label{eqnb47}
& = & s^4_1c^4_2 + 2[c^2_2c^2_3(c^2_1+1) - 2c_1c^2_2c^3_3\cos\delta][c^2_2c^3_3(c^2_1+1)+2c_1c^2_2c^2_3\cos\delta], \\ \label{eqnb48}
& = & s^4_1c^4_2 + 2c^4_2c^4_3[(c^2_1+1) - 2c_1\cos\delta][(c^2_1+1)+2c_1\cos\delta], \\ \label{eqnb49}
& = & s^4_1c^4_2+2c^4_2c^4_3[(c^2_1+1)^2 - 4c^2_1\cos^2\delta], \\ \label{eqnb50}
\langle P_{\nu_\mu\to\nu_\tau}\rangle & = & s^4_1c^4_2 + 2c^4_2c^4_3(c^2_1+1)^2 - 8c^2_1c^4_2c^4_3\cos^2\delta. \label{eqnb52}
\end{eqnarray}
Now according to the topological constraints of Section 3.0 in the main text, (B51), (B42) and (B36) must be 
\emph{equal}. These conditions yield the last constraint on the mixing parameters, namely,
\begin{equation}\label{eqnb53}
c^2_1c^4_2c^4_3\cos^2\delta=0.
\end{equation}
Since $c^2_2=s^2_2=\frac{1}{2}$ and $c^2_3=s^2_3=\frac{1}{2}$ are nonzero, and since it is reasonable to assume [15, 16] that 
the Dirac-type CP-noninvariant phase factor $e^{i\delta}$ is consistent with the requirement that $\sin^2\delta\ne 1$ or $\cos^2\delta\ne 0$ (i.e., CP violation is \emph{not} maximal), we find that $c^2_1=0$ or $s^2_1=1$.  From this point on, the determination of $M$ proceeds exactly as described in the main text (See Eqs.\ 21--28).

\renewcommand \thesection{\arabic{section}.0}
\setcounter{section}{8}
\section*{References and Footnotes}

\noindent[1]  
A.\ P.\ Balachandran, ``Bringing Up a Quantum Baby,'' 
arXiv:quant-ph/9702055.
\medskip

\noindent[2] 
G.\ Holzwarth, ``Formation of Extended Topological-Defects During 
Symmetry-Breaking Phase Transitions in $O(2)$ and $O(3)$ Models'',
arXiv:hep-ph/9901296. Analogous examples of fluctuation-induced topology-change in
macroscopic objects (e.g., destruction of topological objects due to thermal
fluctuations at phase transitions) abounds. For example, otherwise
persistent (``conserved'') \emph{topological defects} in crystals can be
destroyed by raising the temperature sufficiently (i.e., by melting the
crystal). Similarly, otherwise persistent (``conserved'') magnetic
\emph{flux-tubes} in Type II superconductors and/or \emph{vortices} in a
superfluid, can both be destroyed by raising the temperature above the
critical temperature $T_c$. And, conversely, topological defects are always
created when such macroscopic systems first condense (or crystallize) as the
temperature is lowered. We imagine that something roughly similar can happen
when quantum-fluctuations (vacuum fluctuations) act on otherwise very
similar quantum objects (i.e., same electric charge, spin, lepton number, and 
nearly \emph{identical} mass) that also happen
to start ``life'' as topologically-distinct quantum objects. That is, we are assuming
that, if not prevented by some absolute conservation law, transitions between such
states (e.g., $\nu_e$ and $\nu_\tau$ neutrinos) 
will be \emph{catalyzed} by quantum fluctuations.
\medskip

\noindent[3]  
Gerald L. Fitzpatrick, \emph{The Family Problem-New Internal Algebraic
and
Geometric Regularities}, Nova Scientific Press, Issaquah, Washington, 1997.
Additional information:\hfill\break
 {\tt http://physicsweb.org/TIPTOP/} or\hfill\break
{\tt http://www.amazon.com/exec/obidos/ISBN=0965569500}.
\medskip

\noindent[4] 
C.\ Nash and S.\ Sen, \emph{Topology and Geometry for Physicists},
Academic Press, New York, 1983.
\medskip

\noindent[5]
D. J. Thouless, \emph{Topological Quantum Numbers in Nonrelativistic Physics}, World Scientific, Singapore (1998). 
Numerous examples from physics and mathematics could be cited to establish the following very general principle:

\emph{In physical systems characterized by a well defined topology, topology-change tends to be suppressed, relative to the condition of topology-maintenance.}

The basis for this principle is that topology change involves either an energy ``expense'' (an energy ``barrier'' must be overcome) or a violation of some ``topological charge'' conservation law, or both.
In general, we expect this principle to apply to any discontinuous operation such as the \emph{abstract} equivalent of  ``tearing,'' and subsequently ``glueing'' surfaces back together to form states with new topologies. For example, one cannot continuously deform a doughnut into a sphere. It must first be ``cut'' and ``glued'' back together in a new way to achieve such a transformation. 

Now, given the ``facts'' regarding the topology of neutrinos with respect to the (abstract) internal transformation $\bF$,
and \emph{assuming that there are no ``energy barriers'' associated with transitions between neutrinos having the \emph{same} topology}, the foregoing principle suggests that these facts could have the following dynamical significance:

\emph{Topology change in neutrino-neutrino transitions $(\nu_\alpha\to\nu_\beta)$ is suppressed by topological energy and/or topological charge ``barriers,'' while topology maintenance in neutrino-neutrino transitions is relatively enhanced.}

It should be emphasized that even though the $\nu_\mu$ and $\nu_\tau$ neutrinos  have the \emph{same} topology with respect to $\bF$, they are quite \emph{distinct} in other respects. For example, in conventional weak decays or weak capture reactions involving these particles, mu- and tau- numbers are \emph{separately} conserved.
\medskip

\noindent[6] 
The Super-Kamiokande, Kamiokande Collaboration,
arXiv:hep-ex/9810001, and the SNO Collaboration, arXiv:nucl-ex/0106015. While the  matrix proposed to  describe long-distance neutrino mixtures (See Eq.\ 28 in the present paper), is not in conflict with current experimental observations (e.g., $M_{11}=\frac{1}{2}$ is in good agreement with observations of the solar neutrino  deficit), until long-distance neutrino mixtures  (e.g., those from a supernova source) are fully characterized, the status of Eq.\ (28), as a prediction, will not be known.
\smallskip

\noindent[7] 
G.\ L.\ Fitzpatrick, ``Topological Constraints on Long-Distance Neutrino Mixtures,'' \break
arXiv:physics/0007039, 13 July 2000. The proposal for such an equilibrium condition was first put forward in this paper, which is a ``precursor'' to the present, more comprehensive, paper.
\medskip

\noindent[8] 
M.\ Gronau, ``Patterns of Fermion Masses, Mixing Angles and CP Violation,'' in \emph{The Fourth Family of Quarks and Leptons, First International Symposium}, edited by: D.\ B.\ Cline and Amarjit Soni, Annals of The New York Academy of Sciences, New York, New York, Volume 518, 1987, p.\ 190. According to the scheme described in Appendix A of the present paper, the fundamental fermions $d$, $\nu_e(s, b, \nu_\mu, \nu_\tau)$ have the topology of a cylinder (M\"obius strip) with respect to the internal transformation $\bF$. Hence, we imagine that if, in some imaginary world, quark couplings to the Higgs mass-producing fields were nearly ``switched off,'' the resulting ``low'' mass $d$, $s$ and $b$ quark mixtures (located inside quark composites) could conceivably look very much like the $\nu_e, \nu_\mu$ and $\nu_\tau$ low mass neutrino mixtures---owing to topological influences of exactly the kind proposed in the present paper. In particular, in this imaginary world, we would expect the $s$ and $b$ quarks (like the $\nu_\mu$ and $\nu_\tau$ neutrinos, respectively) to exhibit bimaximal mixing. However, because of uncertainty-principle considerations, and because the $b$ quark mass is, in the ``real'' world, very much \emph{larger} than the $d$ and $s$ quark masses, very little mixing with the $b$ quark occurs in $d$, $s$, $b$ quark mixtures \emph{in spite of topological influences that would tend to encourage this}. Thus, in the case of strongly-interacting quarks, it seems very likely that topological constraints of the kind considered in the present paper can, at most, play a \emph{minor} role in determining such things as CKM-type matrix elements. For example, in the case of $d$ and $s$ quarks, Gronau shows that the mixing angle $\theta_c$ depends on quark masses since the matrix element $V_{12}$, and the $d$, $s$ quark masses $m_d$ and $m_s$, respectively, are known to be empirically related via $V_{12}=\sin\theta_c=\sqrt{\frac{m_d}{m_s}}$, where $m_s>m_d\ne 0$. This is nothing like the corresponding matrix element for $\nu_e$ and $\nu_\mu$ neutrinos found in the present paper. Thus, although a detailed understanding of these matters is far from being achieved, it appears that topological influences are allowed to play a major role in neutrino, but \emph{not} in quark mixing, because of the small value, and near degeneracy  of neutrino masses in comparison to the very large differences between $d$, $s$ and $b$ quark masses.
\medskip

\noindent[9]
Although conventional treatments of neutrino mixing do not involve topological considerations, they do describe mixing among neutrinos that initially have a \emph{definite flavor}. And, because we  argue that a neutrino with a definite flavor also has a \emph{definite topology}, hereinafter we will often use these terms together as in ``\dots neutrinos have a definite flavor and topology.''
\medskip

\noindent[10] 
O.\ Nachtmann, \emph{Elementary Particle Physics --- Concepts and Phenomena}, Springer-Verlag, Berlin, 1990, p. 365.
\medskip

\noindent[11] 
Ta-Pei Cheng and Ling-Fong Li, \emph{Gauge Theory of Elementary Particle Physics}, Clarendon Press, Oxford, 1984, pp.\ 409--414. While this reference provides a good summary of the formalism involved in neutrino mixing, the reader should be cautioned that there are several ``typographical'' errors that could cause confusion. For example, in Eq.\ (13.22) the first term in the expression for the matrix element $U_{32}$ should read $c_1s_2c_3$ \emph{not} $c_1s_2s_3$. In Eq.\ (13.31) the last term in the expression for the time-average probability $\lan P_{\nu_e\to\nu_\tau}\ran$ should be preceeded by a \emph{minus} sign, \emph{not} a plus sign.  Finally, the left hand side of Eq.\ (13.33) should read $P_{\nu_\alpha\to\nu_\beta}$ \emph{not} $\lan P_{\nu_\alpha\to\nu_\beta}\ran$.

Note that the time-dependent probability of detecting a neutrino of \emph{definite} flavor $\beta$ at time $t>0$ after emission at time $t=0$ from a source of neutrinos of \emph{definite} flavor $\alpha$, is
\[
P_{\nu_\alpha\to\nu_\beta(t)} = |\lan\nu_\beta(t)|\nu_\alpha\ran|^2,
\]
where the ``scalar product'' $\lan\nu_\beta(t)|\nu_\alpha\ran$ is the ``probability amplitude'' for the process described. For example, if $\alpha=\beta=e$, one has ($\nu_1, \nu_2$ and $\nu_3$ are neutrino mass eigenstates) the ``column'' or ``ket'' vectors
\[
|\nu_e(t)\ran = c_1e^{-iE_1t}|\nu_1\ran+s_1c_3e^{-iE_2t}|\nu_2\ran+s_1s_3e^{-iE_3t}|\nu_3\ran,
\]
and
\[
|\nu_e(0)\ran = |\nu_e\ran=c_1|\nu_1\ran + s_1c_3|\nu_2\ran+s_1s_3|\nu_3\ran.
\]
Then, given  $\lan\nu_i|\nu_j\ran=\delta_{ij}$, and the ``row'' or ``bra'' vector
\begin{eqnarray*}
\lan\nu_e(t)| & = & c_1e^{+iE_1t}\lan\nu_1|+s_1c_3e^{+iE_2t}\lan\nu_2|+s_1s_3e^{+iE_3t}\lan\nu_3|, \\
P_{\nu_e\to\nu_e(t)} & = & |\lan \nu_e(t)|\nu_e\ran|^2 \\
& = & |(c^2_1\cos E_1t+s^2_1c^2_3\cos E_2t+s^2_1s^2_3\cos E_3t)\\
&&\qquad + i(c^2_1\sin E_1t+s^2_1c^2_3\sin E_2t+s^2_1s^2_3\sin E_3t)|^2 \\
& = & (c^2_1\cos E_1t+s^2_1c^2_3\cos E_2t+s^2_1s^2_3\cos E_3t)^2 \\
&&\qquad + (c^2_1\sin E_1t + s^2_1c^2_3\sin E_2t+s^2_1s^2_3\sin E_3
t)^2.
\end{eqnarray*}

After expansion, and further simplification, the previous expression becomes [use $\cos(A-B)=\cos A\cos B+\sin A\sin B$]
\begin{eqnarray*}
P_{\nu_e\to\nu_e(t)} = (1-2c^2_1s^2_1-2s^4_1c^2_3s^2_3) & + & 2c^2_1s^2_1c^2_3\cos(E_1-E_2)t \\
& + & 2c^2_1s^2_1s^2_3\cos(E_1-E_3)t \\
& + & 2s^4_1c^2_3s^2_3\cos(E_2-E_3)t.
\end{eqnarray*}
In general, at time $t$ or distance $X$ from the neutrino source, one has the expression
\[
P_{\nu_\alpha\to\nu_\beta(t)} = \sum^3_{i=1}|U_{\alpha i}|^2|U_{\beta i}|^2 + \sum^3_{i\ne j} U_{\alpha i}U^\ast_{\beta i} 
U^\ast_{\alpha j}U_{\beta j}\cos\left(\frac{2\pi X}{l_{ij}}\right).
\]
where $l_{ij}=2\pi/(E_i-E_j)$, and $E_i-E_j=(m^2_i-m^2_j)/2p$.

Clearly, the general time-average probability, when $X\gg l_{ij}$ is given by (note that time-averages of each of the three cosine terms above produce ``sinc'' functions varying as $\frac{\sin kX}{kX}$, each of which approaches \emph{zero} as the magnitude of $X$ increases) 
\[
\lan P_{\nu_\alpha\to\nu_\beta(t)}\ran = \sum^3_{i=1} |U_{\alpha_i}|^2|U_{\beta_i}|^2.
\]
For example, referring to the expression for $P_{\nu_e\to\nu_e(t)}$ worked out above, one sees that 
\[
\lan P_{\nu_e\to\nu_e(t)}\ran = \sum^3_{i=1}|U_{e_i}|^2|U_{e_i}|^2,
\]
or
\[
\lan P_{\nu_e\to\nu_e(t)}\ran = (1-2c^2_1s^2_1-2s^4_1c^2_3s^2_3).
\]
Note that $P_{\nu_\alpha\to\nu_\beta(t)}$ and $\lan P_{\nu_\alpha\to\nu_\beta(t)}\ran$ are also commonly written as $P_{\nu_\alpha\to\nu_\beta}$ and $\lan P_{\nu_\alpha\to\nu_\beta}\ran$, respectively.
\medskip

\noindent[12] 
H.\ Fritzsch and Z.\ Xing, ``How to Describe Neutrino  Mixing and CP Violation,'' arXiv:hep--ph/0103242. CP- and T-violating asymmetries in ``normal'' (lepton-number conserving)  neutrino oscillations depend only on the Dirac-type phase $\delta$. In particular, they have nothing to do with the Majorana-type CP-violating phases $\rho$ and $\sigma$. Moreover, because lepton-number violating processes (e.g., neutrinoless double beta decay) are known from experiment to be strongly suppressed,  the issue of whether neutrinos are Dirac particles or Majorana particles need not be addressed in the present paper. However, it should be understood that if neutrinos are Majorana particles, lepton-number violating processes would be possible (if not probable!), and the matrix $U_{\alpha i}$ in Eqs.\ (1) and (2) in the main text would have to be multiplied by a diagonal matrix such as
\[
\left(
\begin{array}{ccc}
1 & 0 & 0 \\
0 & e^{i\rho} & 0 \\
0 & 0 & e^{i\sigma} \end{array}\right), 
\]
where $\rho$ and $\sigma$ are phases associated with the Majorana-type CP-violating  phase factors $e^{i\rho}$ and $e^{i\sigma}$, respectively.
\medskip

\noindent[13]
By definition, mixing does not occur at the source of neutrinos. Hence, Eq.\ (4) in the main text is obviously true.  For example, at the source, neutrinos are \emph{pure} flavor eigenstates $\nu_\alpha$, hence
\[
P_{\nu_\alpha\to\nu_\beta}\Bigg|_{X=0} = |\lan\nu_\beta|\nu_\alpha\ran|^2=|\delta_{\beta\alpha}|^2,
\]
where $\delta_{\beta\alpha}=1$ when $\alpha=\beta$, and $\delta_{\beta\alpha}=0$ when $\alpha\ne\beta$.
However, it is instructive to demonstrate the validity of Eq.\ (4) for at least one matrix element, starting with Eq.\ (3)
in the main text.
In particular, let us demonstrate that when $X=0$, $P_{\nu_e\to\nu_e}=1$. The other matrix elements can be found in a similar way, which demonstrates the validity of Eq.\ (4).

When $X=0$, Eq.\ (3) becomes (Use Eq.\ 5)
\begin{eqnarray*}
P_\nue & = & \lan P_\nue\ran + \sum^3_{i\ne j} U_{1i}U^\ast_{1i}U^\ast_{1j}U_{1j} \\
& = & \lan P_\nue\ran + \begin{array}[t]{l}
U_{11}U^\ast_{11}U^\ast_{12}U_{12} + U_{11}U^\ast_{11}U^\ast_{13}U_{13} + \\
U_{12}U^\ast_{12}U^\ast_{11}U_{11} + U_{12}U^\ast_{12}U^\ast_{13}U_{13} + \\
U_{13}U^\ast_{13}U^\ast_{11}U_{11} + U_{13}U^\ast_{13}U^\ast_{12}U_{12}\end{array} \\
& = & \lan P_\nue\ran \begin{array}[t]{l}
+|U_{11}|^2|U_{12}|^2 + |U_{11}|^2|U_{13}|^2 + |U_{12}|^2|U_{11}|^2 \\
+ |U_{12}|^2|U_{13}|^2+|U_{13}|^2|U_{11}|^2 + |U_{13}|^2|U_{12}|^2\end{array} \\
& = & \lan P_\nue\ran \begin{array}[t]{l}
+|U_{11}|^2\{|U_{12}|^2+|U_{13}|^2+|U_{12}|^2+|U_{13}|^2\} \\
+ |U_{12}|^2\{|U_{13}|^2+|U_{13}|^2\}\end{array}\\
P_\nue & = & \lan P_{\nue}\ran + 2|U_{11}|^2\{|U_{12}|^2+|U_{13}|^2\} + 2|U_{12}|^2|U_{13}|^2.
\end{eqnarray*}
Given (B6) and Eq.\ (2) in the present paper, one has
\[
P_\nue=1-2c^2_1s^2_1-2s^4_1s^2_3c^2_3+2c^2_1s^2_1\{c^2_3+s^2_3\}+2s_1^4s^2_3c^2_3
\]
or
$P_\nue=1$. Proceeding in this way with the calculation of the other matrix elements, one establishes the intuitively 
obvious result
\[
P_{\nu_\alpha\to\nu_\beta}\Bigg|_{X=0}=I_3.
\]
\medskip

\noindent[14]
The proposed topological constraints are of course \emph{assumed}. We imagine that they are to be justified by physics at some deeper level than the conventional description of neutrino mixing. While we have no way to prove these \emph{qualitative} assumptions regarding topological constraints, their adoption certainly leads to testable predictions regarding mixing parameters, and the matrix of time-average probabilities $\langle P_{\nu_\alpha\to\nu_\beta}\rangle$.
\medskip

\noindent[15]
E.\ Rodriguez-Jauregui, ``Implications of Maximal Jarlskog Invariant and Maximal CP-Violation,'' arXiv: hep-ph/0104092. While the author of this reference argues that \emph{maximal} CP-violation may occur in both the quark and lepton sectors, these arguments do not seem to be compelling. Instead, I find it necessary in the present paper to assume the contrary, namely, that CP-violation is \emph{not} maximal, at least in the lepton sector.

According to Eqs.\ (20) and (B52) in the present paper, the topological constraints of Section 3.0 have forged a kind of ``linkage'' between the mixing angle $\theta_1$, and the angle
$\delta$ associated with the Dirac-type CP-noninvariant phase factor $e^{i\delta}$, namely,
\[
c^2_1\cos^2\delta=0.
\]
Even though we do not know the physical origin of CP-noninvariance, it is clear that if we make what seems to be a reasonable assumption, namely,  that CP-violation is \emph{not} maximal (e.g., $\delta$ is \emph{not} equal to $\pm\frac{\pi}{2}$), $\delta$ will be subject to the constraint $\sin^2\delta\ne 1$ or  $\cos^2\delta\ne 0$, in which case $\theta_1$ will be constrained by $c^2_1= 0$ or $c_1= 0$. 

In this case, Eq.\ (2) yields, for example (Use $s_1=1$)
\[
\nu_e=s_1c_3\;\nu_2+s_1s_3\;\nu_3.
\]
But, the topological constraints of Section 3.0 have also yielded the constraint $c^2_3=s^2_3 =\frac{1}{2}$, which means that (For the sake of argument assume $\theta_3$ is a first-quadrant angle)
\[
\nu_e=\frac{\sqrt{2}}{2}\nu_2+\frac{\sqrt{2}}{2}\nu_3.
\]
Clearly, this last equation describes ``bi-maximal'' mixing of the \emph{mass} eigenstates $\nu_2$ and $\nu_3$.
\medskip

\noindent[16] 
H.\ Georgi and S.\ L.\ Glashow, ``Neutrinos on Earth and in the
Heavens,'' arXiv:hep-ph/9808293, and hep-ph/9808293v2, page 5, Eq.\ (20). It is interesting, and probably significant that Georgi and Glashow  independently arrived at Eq.\ (28) in the main text of the present paper starting from six experimental neutrino ``facts,'' some of which are completely different than the facts and assumptions employed in the present paper. In particular, these authors began by assuming
\begin{enumerate}
\item There are just three chiral neutrino states having Majorana masses.
\item Atmospheric neutrinos rarely oscillate into electron neutrinos.
\item Atmospheric muon neutrinos suffer maximal, or nearly maximal, two flavor oscillations into tau neutrinos.
\item[4. \& 5.]Two experimentally (and theoretically) motivated assumptions regarding the order of magnitude of individual neutrino masses, and various mass squared differences.
\end{enumerate}
and
\begin{enumerate}
\item[6.]Neutrinoless double beta decay probably does \emph{not} occur.
\end{enumerate}
Given these six ``facts'' these authors derived both the neutrino mass matrix, and partially determined the associated CKM-like unitary matrix $U$, which describes neutrino mixtures. From $U$ they then derived Eq.\ (28) in the present paper. In closing, these authors also noted that CP-violation in this situation could be ``superweak.'' Clearly, this would be consistent with the assumption in the present paper that CP-violation is \emph{not} maximal, i.e., $\sin^2\delta\ne 1$ or $\cos^2\delta\ne 0$.

While the present approach, and that of Georgi and Glashow in deriving Eq.\ (28), are quite different they are, nevertheless, expected to be compatible. In particular, we anticipate that such things as \emph{neutrino masses and mixing parameters, ultimately owe their existence to physics at a deeper level, where topological considerations, of the kind proposed in the present paper,  are expected to play an important role.}
\medskip

\noindent[17]
Note that for all practical purposes, time-dependent \emph{quantum} probabilities $P_{\nu_\alpha\to\nu_\beta}$ become time-average or \emph{classical} probabilities $\lan P_{\nu_\alpha\to\nu_\beta}\ran$ when $t$ is sufficiently large. For this reason, it makes sense to consider classical probability measures such as the ``joint'' probability $P=P_CP_{NC}$, where $P_C$ and $P_{NC}$ are time-averages of quantum probabilities.
\medskip

\noindent[18]
Even if the matrix $M$ in Eq.\ (28) in the main text is verified by experiment, we could not be certain that the constraints expressed by Eqs.\  (9), (12) and (13), together with items A), B) and C) in Section 3.0, are \emph{topological} constraints related to the 2-space description of quarks and leptons. There are three basic reasons for this assertion.

First, we do not know why the topology of vector triads (with respect to the internal transformation $\bF$) should be relevant to neutrino mixing. Second, we do not have any detailed understanding of the hypothetical mechanism that supplies the ``energy barriers'' or ``topological charge conservation laws'' that serve to inhibit topology change in neutrino mixing. Third, as indicated in Appendix A (Sec.\ A.4), we cannot be absolutely certain that the $\nu_\mu$ neutrino  has the requisite second-family 
M\"obius topology with respect to $\bF$. 

What we do know with certainty is that, regardless of their physical origins, the  constraints of Section 3.0, together with the additional assumption $\sin^2\delta\ne 1$ or $\cos^2\delta\ne 0$, definitely determine $M$ \emph{uniquely}. 
\medskip

\noindent[19] 
G.\ L.\ Fitzpatrick, ``Continuation of the Fermion-Number Operator and the Puzzle of Families,'' arXiv:physics/0007038, 13 July 2000.
\medskip

\noindent[20]
G.\ L.\ Fitzpatrick, ``Electric Charge as a Vector Quantity,'' arXiv:physics/0011073, 30 November 2000.
\medskip

\noindent[21]
S.\ Weinberg, \emph{The Quantum Theory of Fields, Vol.\ I, Foundations}, Cambridge University Press, New York, NY (1995), pp.\ 529--531; \emph{The Quantum Theory of Fields, Vol.\ II, Modern Applications}, Cambridge University Press, New York, NY (1996), p.\ 155.

\end{document}